\documentclass[doublecol]{epl2} 
\usepackage{subfigure}
\usepackage{amsmath}
\usepackage{color}
\usepackage{wrapfig}
\usepackage{cite}

\newcommand{\pdiff}[2]{\frac{\partial #1}{\partial #2}}

\newcommand{\be}{\begin{equation}}
\newcommand{\ee}{\end{equation}}
\newcommand{\ba}{\begin{eqnarray*}}
\newcommand{\ea}{\end{eqnarray*}}

\newcommand{\abs}[1]{\left|#1\right|}

\newcommand{\eper}{\varepsilon_{\perp}}
\newcommand{\epar}{\varepsilon_{\parallel}}

\title{Fredrickson-Andersen model on Bethe lattice with random pinning}
\shorttitle{Fredrickson-Andersen model on Bethe lattice with random pinning} 

\author{Harukuni Ikeda\inst{1} \and Kunimasa Miyazaki\inst{1}}
\shortauthor{H. Ikeda and K. Miyazaki}

\institute{                    
  \inst{1} Department of Physics, Nagoya University, Nagoya 464-8602, Japan
}

\abstract{ We study the effects of random pinning on the
Fredrickson-Andersen model on the Bethe lattice.  We find that the
nonergodic transition temperature rises as the fraction of the pinned
spins increases and the transition line terminates at a critical point.
The freezing behavior of the spins is analogous to that of a randomly
pinned $p$-spin mean-field spin glass model which has been recently
reported. The diverging behavior of correlation lengths in the vicinity
of the terminal critical point is found to be identical to the
prediction of the inhomogeneous mode-coupling theory at the $A_3$
singularity point for the glass transition.}

\begin{document}

\maketitle
\section{Introduction}
The nature of the glass transition is still
elusive~\cite{Debenedetti2001,biroli2013perspective,gotze2008complex}.
It is a major challenge to understand universal behavior of the
transition, such as the non-Arrhenius growth of the viscosity,
non-exponential relaxation of the correlation functions, and spatially
heterogeneous dynamics of the fluctuations. The ultimate goal is to
establish whether the transition is associated with any thermodynamic
singularity, or it is caused by purely kinetic
mechanism~\cite{Debenedetti2001,biroli2013perspective}. Obvious reasons
that hamper the progress of our understanding of the glass transition
are the inhibitedly long time scales required to reach the transition
point experimentally and the lack of an ideally simple
finite-dimensional model which exhibits a true glass transition, if any.

Recently, a novel idea to bypass the difficulty to access the glass
transition temperature by randomly freezing, or pinning, a fraction of
degrees of freedom of the equilibrated system has been
proposed~\cite{kim2003effects,PhysRevLett.94.065703,cammarota2012ideal,cammarota2013random}.
Cammarota {\em et al.} have analyzed the effects of random pinning on
the glass transition of the $p$-spin mean-field spin glass
model\cite{cammarota2012ideal,cammarota2013random}. It was found that
both the ideal glass transition temperature, $T_K$, and the dynamic
transition temperature, $T_d$, rise as the fraction of the pinned spins,
$c$, increases. Furthermore, the two transition lines, $T_K(c)$ and
$T_d(c)$, are found to merge and terminate at a finite $c$. This end
point is argued to be a critical point whose universality class is that
of the mean-field random-field Ising model\cite{biroli2014random}.  In
the terminology of the mode-coupling theory (MCT), which is a mean-field
dynamical theory of the glass transition, this end point is
characterized as the $A_3$ singularity, where the anomalous dynamical
scalings, such as the logarithmic relaxation dynamics and distinct
diverging length scales, are
predicted~\cite{nandi2014critical}. Verification of the ideal glass
transition of randomly pinned systems by experiments and simulations for
realistic systems may be crucial to prove (or disprove) the very
existence of the bona-fide glass transition point at finite dimensions
\cite{PhysRevLett.110.245702,ozawa2014equilibrium}. At the same time, it
is imminent to establish the relationship of thermodynamic scenarios
such as the random first order transition theory (RFOT) with other
kinetic scenarios which do not necessarily require thermodynamic
singularities behind the glassy slow
dynamics~\cite{kirkpatrick1987p,kirkpatrick1989scaling,ritort2003glassy,biroli2013perspective}.

In this letter, we argue that the singular behavior of the glass
transition by random pinning analogous to that of the $p$-spin
mean-field spin glass model can be also observed for a purely kinetic
model on the Bethe lattices (or random graphs).  Kinetically constrained
models (KCMs) are toy lattice or spin models in which Hamiltonian is
free from the interaction but a non-trivial constraint is imposed on the
dynamic rule~%
\cite{PhysRevLett.53.1244,ritort2003glassy,biroli2013perspective}. Many
KCMs are known to display slow dynamics in the collective movement of
the particles or spins at low temperatures but they are of purely
kinetic origin since their thermodynamics are trivially ideal-gas like.
Their glassy behavior is genetically distinct from those of
thermodynamic scenarios such as RFOT where thermodynamic singularities
encoded in the free energy landscape play a pivotal role and escort the
slow dynamics.  Indeed, most KCMs show no glass transition at finite
temperatures.  Notable exceptions are KCMs on the Bethe lattices (or
random graphs)\cite{ritort2003glassy,sellitto2005facilitated}.  One of
the KCMs called the Fredrickson-Andersen model (FAM) is known to
undergo a nonergodic transition called the Bootstrap Percolation (BP)
transition at a finite temperature $T_d$ on the Bethe
lattice~\cite{PhysRevLett.53.1244,chalupa1979bootstrap,sellitto2005facilitated}.
It has been demonstrated that the time evolution of the order parameter
of the FAM shows dynamical anomaly similar to those predicted
by the MCT for structural glass formers, such as the two step relaxation
and algebraic divergence of the relaxation time near
$T_d$~\cite{sellitto2005facilitated}.
Here, we demonstrate that the randomly pinned
FAM on the Bethe lattice also shows the analogous critical
behavior as that of the randomly pinned $p$-spin mean-field spin glass
model such as the $A_3$ higher order
singularity\cite{cammarota2012ideal,cammarota2013random}.

\section{Model}

We consider a FAM on the Bethe lattice. In the absence of pinned spins,
this model is defined as follows. We consider $N$ Ising spins
$\sigma_i\in\{-1,+1\}$, $i=1,\cdots, N$.  The Hamiltonian of the system
is free from the interaction and can be written as
\begin{equation}
H=-\frac{1}{2}\sum_{i=1}^{N}\sigma_i.
\end{equation}
The equilibrium distribution of the up spins is given by
\begin{equation}
p=\frac{1}{1+\exp(-1/T)},\label{111919_19Mar15}
\end{equation}
where $p$ varies from $1/2$ to 1, corresponding to $T=\infty$ and $T=0$,
respectively.

The spin variables represent coarse grained mobility fields.
$\sigma_i=+1$ represents the immobile region and $\sigma_i=-1$ is the
mobile region or defect~\cite{ritort2003glassy}. The idea of the FAM is
to introduce a kinetic constraint on the time evolution of the spins in
such a way that flipping of a spin is more difficult when it is
surrounded by many up spins (immobile regions).  To be more specific, the
$i$-th spin at each Monte-Carlo step can flip with a transition
probability
\begin{equation}
w(\sigma_i\to -\sigma_i)={\rm min}\{1,e^{-\sigma_i/T}\}, 
\label{eq:transition-probability}
\end{equation}
only if the number of the nearest spins in the state $-1$ is larger than
or equal to $f$~\cite{PhysRevLett.53.1244,ritort2003glassy}.  It is
known that the FAM on a lattice in finite dimensions does not exhibit
the transition at a finite temperature but it does so on the Bethe
lattices and random graphs~\cite{ritort2003glassy}.  The FAM undergoes
the nonergodic transition at a finite temperature if the connectivity
$k$, or the number of the nearest neighbours $k+1$, satisfies
$k>f>1$~\cite{sellitto2005facilitated}.

\begin{figure}
 \onefigure[width=4.5cm]{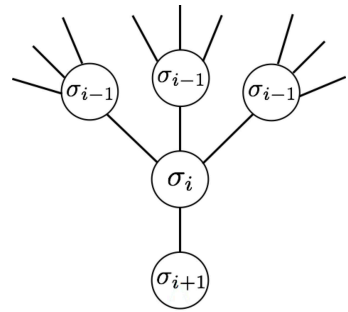}
\caption{\small{The Bethe lattice with $k=3$.}} 
\label{fig1}
\end{figure}

Now we consider to pin or freeze a fraction $c$ of the spins randomly
chosen from the $N$ spins which are initially equilibrated at a
temperature $T$ before pinning.  Note that pinning spins in the
equilibrated configuration at $T$ is essential in the following
argument. The dynamic rule after pinning is basically the same as the
bulk system; the $i$-th spin can flip with the probability of
eq.(\ref{eq:transition-probability}), only if the number of the nearest
spins in the state $-1$ is larger than or equal to $f$ {\em and the
$i$-th spin is not pinned}.  In the following, we only consider the case
of $k=3$ and $f=2$ (see Fig.~\ref{fig1}), but conclusions for different
sets of $(k,f)$ do not change qualitatively as long as $k > f > 1$.

\section{Phase diagram}

Long-time behavior of the randomly pinned FAM on the Bethe lattice
starting from the equilibrium distribution at the initial time can be
evaluated analytically invoking the recursive relation on the lattice.
We shall evaluate the persistent function $\phi$, which is the total
fraction of the dynamically frozen spins at the long-time limit, defined
by
\begin{equation}
 \phi =  c+ P_{+} + P_{-}.\label{080259_27Jan15}
\end{equation}
Here the first term $c$ represents the fraction of the pinned spins and
$P_{\pm}$ represent the fraction, or the probability, of the spins
ultimately arrested in the $+(-)$ state among the $(1-c)N$ unpinned spins,
respectively.  $P_{+}$ is written as
\begin{equation}
 P_{+} = (1-c)p  \sum_{n=0}^{f-1}\binom{k+1}{n}B^{n}(1-B)^{k+1-n},
\label{eq:Q-}
\end{equation}
where $p$ is given by eq.(\ref{111919_19Mar15}) and $B$ is the
conditional probability that a spin is in the state $-1$ or flipped down
to the state $-1$ given that one of the nearest neighbours was arrested
in the state $+1$. $B$ obeys the following expression;
\begin{equation}
 \begin{aligned}
B &= 1-p+(1-c)p \sum_{n=0}^{k-f}\binom{k}{n}B^{k-n}(1-B)^{n}.
 \end{aligned}
\label{eq1} 
\end{equation}
The first term, $1-p$, on the right hand side of eq.(\ref{eq1})
represents the equilibrium probability of $-1$ spins.  The fact that it
is independent of $c$ is the reflection that the spins have been
randomly pinned from the equilibrium distribution. The second term
represents the probability of the spins originally at the state $+1$ but
eventually flipped down to the state $-1$.  Eq.~(\ref{eq1})
is closely related to that of the multi-component extension of the FAM
and associated models in Refs
\cite{PhysRevLett.105.265704,branco1993probabilistic}.  Indeed,
eq.(\ref{eq1}) can be obtained by freezing one of the degree of freedom in
the equation for the binary system considered in Ref
\cite{branco1993probabilistic} (see eq.(3) therein).  Likewise,
$P_{-}$ can be written as
\begin{equation}
 P_{-} = (1-c)(1-p)\sum_{n=0}^{f-1}\binom{k+1}{n}B^{\prime n}(1-B^{\prime})^{k+1-n},
\label{eq:Q+}
\end{equation}
where $B^{\prime}$ is the conditional probability that a spin is in the
state $-1$ or flipped down to the state $-1$ given that one of the
nearest neighbours was arrested in the state $-1$, which is given by the
solution of
\begin{equation}
B' = 1-p  + (1-c)p \sum_{n=0}^{k-f+1}\binom{k}{n}B^{k-n}(1-B)^{n}. 
\label{040411_25Dec14}
\end{equation}
From eqs.(\ref{eq:Q-}) and (\ref{eq:Q+}), the total fraction of the
arrested spins, eq.(\ref{080259_27Jan15}), is written as
\begin{eqnarray}
\phi = c+(1-c)
\left[
p \Psi_{k+1}^{f}(B) 
+
(1-p)
\Psi_{k+1}^{f}(B^{\prime}) 
\right],
\label{phi}
\end{eqnarray}
where we have defined an auxiliary function 
\begin{equation}
\Psi_{k}^{f}(B) \equiv  
\sum_{n=0}^{f-1}\binom{k}{n}(1-B)^{k-n}B^{n}. 
\end{equation}
One observes that $B'$ and $\phi$ can be computed from the
self-consistent equation for $B$ given by eq.(\ref{eq1}).  In the limit
of $c=0$, these expressions reduce to those studied in
Ref.\cite{sellitto2005facilitated}.  Eq.~(\ref{phi})
obtained for the stationary state coincides with the long-time limit of
the time-dependent persistent function $\phi(t)$ evaluated for the
equilibrium initial configuration\cite{ritort2003glassy,sellitto2005facilitated,PhysRevLett.105.265704,arenzon2012microscopic}.
We checked this by the Monte-Carlo (MC) simulation for several state points
(see the inset of Fig.~\ref{fig3}).
\begin{figure}
\onefigure[width=8.8cm]{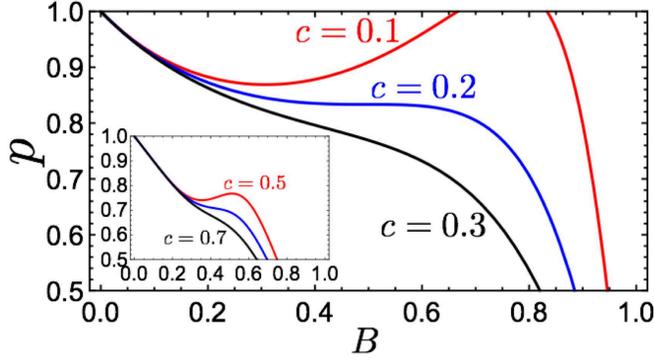} 
 \vspace{-0.5cm}
\caption{\small{ $p$ as a function of $B$ for $k=3$ and
$f=2$. Three lines correspond to $c=0.1$, $0.2$ and $0.3$ from the top
to the bottom. Inset: Same results for $k=15$ and $f=7$, where $c=0.5$,
$0.6$ and $0.7$ from the top to the bottom.}}  \label{fig2}
\end{figure}
For $k=3$ and $f=2$, eq.(\ref{eq1}) can be explicitly written as 
\begin{equation}
B = 1-p + (1-c)p \left\{ B^3 + 3B^{2}(1-B) \right\}.
\label{eq2}
\end{equation}
If $c=0$, there is always a trivial solution $B=1$, {\em
i.e.}, all spins can flip eventually and the system is ergodic.  However, when
the temperature is lowered from above, or $p$ is raised from below from
$p=1/2$, $B$ discontinuously jumps from $1$ to $B_d =1/4$ at a transition
point, $p_d=8/9$, and concomitantly $\phi$ jumps from $0$ to a finite
value of $\phi_d=0.673$. This is the $A_2$ transition in the MCT
terminology~\cite{gotze2008complex}.

If $c\neq 0$, $B\neq 1$ always holds and eq.(\ref{eq2}) can be solved
immediately to obtain
\begin{equation}
p(B,c) = \frac{1-B}{1-(1-c)\left[ B^3+3B^2(1-B) \right]}.\label{121937_19Mar15}
\end{equation}
In Fig.~\ref{fig2}, we show $p(B,c)$ as a function of $B$ for several
$c$'s.  If $c\neq 0$ but small, $B$-dependence of $p$ shows a
nonmonotonic behavior (see $c=0.1$ of Fig.~\ref{fig2}).  When $p$ is
small (high temperature) where most of the spins are down (or mobile),
there is only one solution for $B$ close to $1$, where a small fraction
of the spins freezes in the vicinity of the pinned spins and most of the
spins can still flip.  As we increase $p$, the nontrivial branches at a
small $B$ (or a large $\phi$) appear discontinuously at $p=p_d$, which
is the signal of the $A_2$ singularity.
\begin{figure}
\onefigure[width=8.8cm]{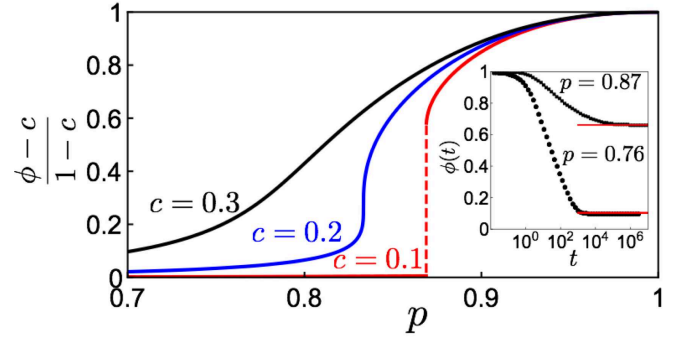}
 \vspace{-0.5cm}
 \caption{\small{$p$ dependence of
 $(\phi(c,p)-c)/(1-c)$ for several $c$'s: The transition is
 discontinuous at $c<1/5$ and continuous at $c=1/5$. For $c>1/5$, only
 the crossover is observed. Inset: The $t$-dependence of $\phi(t)$ for $c=0.1$
 and several $p$'s obtained by MC simulation (filled symbols). The solid lines are the
 solutions of the stationary equation, eq.~(\ref{phi}).}}  \label{fig3}
\end{figure}
The width of the gap of the discontinuous jump of $B$ decreases as $c$
is increased. As $c$ increases further, the situation qualitatively
changes. The minimum of $p(B, c)$ at $B_d$ becomes unstable and
eventually disappears. It is clear from eq.~(\ref{121937_19Mar15}) that
this happens at $c_s=1/5$.  At this point, the $A_2$ singularity
disappears and the transition becomes continuous. This is the signal of
the $A_3$ singularity. The MCT predicts that the critical behavior of
the $A_3$ singularity is distinct from that of the $A_2$
singularity~\cite{gotze2008complex,nandi2014critical,cammarota2013random,cammarota2012ideal}.
Qualitatively the same trend is observed for different sets of $k$ and $f$
(see the inset of Fig.~\ref{fig2}).

$\phi(c, p)$ evaluated by numerically solving
eqs.~(\ref{040411_25Dec14})$-$(\ref{eq2}) is plotted in Fig.~\ref{fig3}
for several $c$'s.  The behavior of $\phi(c,p)$ is qualitatively the
same as that of the nonergodic parameter evaluated from the MCT across
the $A_3$ singularity point.  Note that similar $A_3$ singularities have
been analyzed for binary and ternary mixtures of the FAM on the Bethe
lattices and random graphs
\cite{branco1993probabilistic,PhysRevLett.105.265704,arenzon2012microscopic,cellai2013critical}.
\begin{figure}
 \onefigure[width=7cm]{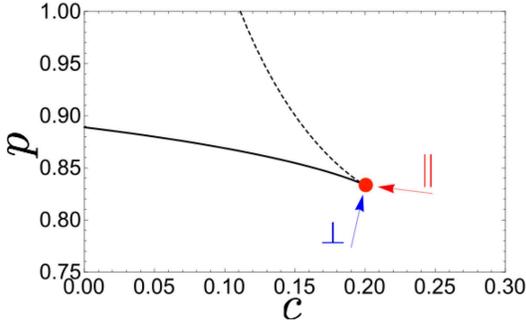} \caption{\small{The (dynamic) phase
diagram of the randomly pinned FAM: The thick solid line is the
nonergodic transition line $p_d(c)$.  The dashed line is a ``spinodal''
line at which the larger branches of $B$ disappear.  The filled circle at
the end of the thick solid line at $(c=1/5,p=5/6)$ is the terminal
critical point of the higher order singularity. The arrows represent the
directions of vectors $\vec{e}_{\perp}$ and $\vec{e}_{\parallel}$ (see
the text).} } \label{fig4}
\end{figure} 

The nonergodic transition line as a function of $c$, $p_d(c)$, is shown
in Fig.~\ref{fig4} in the thick solid line.  Recall that there exist
three solutions of $B$ of eq.~(\ref{eq2}) if $p$ is slightly above $p_d$
and the smallest $B$ is the real solution.  As it is observed in
Fig.~\ref{fig2}, if $c$ is slightly smaller than $c_s$, there is always
a finite $p$ $(>p_d)$ above which the larger solutions of $B$ disappear.
In Fig.~\ref{fig4}, this ``spinodal'' line is shown in dashed line.
This ``phase-diagram'' demonstrates that the transition point $p_d$
decreases (or $T_d$ increases) as $c$ increases and eventually
terminates at a critical point. The critical point can be evaluated
analytically as $c_s=1/5$ and $p_s=p_d(c_s)=5/6$.  At $c>c_s$, there is
no distinction between the ergodic (fluid) and nonergodic (glassy)
phases and $\phi$ continuously increases as $p$ increases.  The behavior
of $p_d(c)$ in Fig.~\ref{fig4} is qualitatively the same as that
evaluated for the randomly pinned $p$-spin mean-field spin glass
model~\cite{cammarota2012ideal,cammarota2013random}.  We
present the phase diagram for $k=3$ and $f=2$ but the results remain
unchanged for other sets of $k$ and $f$, as long as $k>f>1$.

\section{Higher Order Singularity}
If $c<c_s$, the transition across the thick solid line in
Fig.~\ref{fig4} is discontinuous and the order parameters, $\phi$ (or
$B$), behave as $\phi-\phi_d$ (or $B-B_d$) $\propto \abs{p_d-p}^{1/2}$
\cite{sellitto2005facilitated}. At $c=c_s=1/5$, however, the properties
of the transition qualitatively change. In the following, we show that
the critical behavior around this terminal critical point is indeed
characterized as the $A_3$
singularity~\cite{gotze2008complex,nandi2014critical,
sellitto2013disconnected,sellitto2012cooperative}.

We first introduce a function defined by
\begin{equation}
Q(c,p,B) \equiv 1-p + (1-c)p \left\{B^3+3B^2(1-B)\right\}-B
\label{eq:Q-def}.
\end{equation}
From eq.(\ref{eq2}), $Q=0$ always holds.
We shall expand this function around the
transition point as
\begin{equation}
\begin{aligned}
Q(c,p,B) 
= & 
\pdiff{Q}{B}\delta B
 + \frac{1}{2}\pdiff{^2Q}{B^2}\delta B^2 
+ \frac{1}{3!}\pdiff{^3Q}{B^3}\delta B^{3}
+ \pdiff{Q}{c}\delta c\\
&+ \pdiff{Q}{p}\delta p
+ \left( \pdiff{^{2}Q}{c\partial B}\delta c +\pdiff{^2Q}{p\partial B}\delta p \right)\delta B
+ \cdots  \label{eq:Q},
\end{aligned}
\end{equation}
where $\delta B=B-B_d$ and $\delta c = c-c_d$.
The transition point is called the $A_2$ singularity point if $\partial
Q/\partial B=0$ and $\partial^2 Q/\partial B^2\neq 0$, and the $A_3$
singularity point if $\partial Q/\partial B=\partial^2Q/\partial B^2=0$ and
$\partial^3Q/\partial B^3\neq
0$~\cite{nandi2014critical,sellitto2013disconnected,sellitto2012cooperative}.  It is obvious
that on the transition line, $p_d(c)$, of Fig.~\ref{fig4} for small $c$,
$\partial Q/\partial B=0$ and $\partial^2 Q/\partial B^2\neq 0$ are
satisfied and that they are of the $A_2$ type. However,
exactly on the terminal critical point, $(c_s,p_s,B_s)=(1/5,5/6,1/2)$,
$\partial^2Q/\partial B^2$ vanishes and this point is categorized as
the $A_3$ singularity point. In the vicinity of the terminal critical point,
eq.(\ref{eq:Q-def}) can be written as
\begin{eqnarray}
0\approx  \frac{1}{12}\varepsilon_c
+ \frac{1}{2}\varepsilon_p + \left( \frac{3}{16}\varepsilon_c
 + \frac{3}{4}\varepsilon_p
 \right)\delta B -\frac{4}{3}\delta B^{3},
 \label{eq: Qexpansion}
\end{eqnarray}
where we defined $\varepsilon_c=(c_s-c)/c_s$ and
$\varepsilon_p=(p_s-p)/p_s$.  Note that we have two control parameters,
$\vec{\varepsilon}=(\varepsilon_c, \varepsilon_p)$.  In order to
investigate the critical behavior, it is convenient to transform the set
of control parameters parallel and perpendicular to the transition
line. Since the first two terms of eq.~(\ref{eq: Qexpansion}) are
written in the form of the product of $\vec{\varepsilon}$ and a constant
vector $(1/12,1/2)$, it is natural to introduce the unit vectors
directing to and perpendicular to this constant vector by
$\vec{e}_{\perp} = \frac{1}{\sqrt{37}} \left(1,6\right)$ and
$\vec{e}_{\parallel} = \frac{1}{\sqrt{37}} \left(-6,1\right)$ (see
arrows in Fig.~\ref{fig4}), so that
 $\vec{\varepsilon}$ can be written as
$\vec\varepsilon = \eper\vec{e}_{\perp}+\epar\vec{e}_{\parallel}$. With
this expression, eq.(\ref{eq: Qexpansion}) can be rewritten as
\begin{eqnarray}
0=-\frac{4 \delta B^3}{3}+\frac{\sqrt{37} \eper}{12} 
+\frac{3\delta B\left(25\eper-2\epar\right)}{16\sqrt{37}}
, \label{011155_26Dec14}
\end{eqnarray}
and one immediately finds $\delta B \approx\abs{\eper}^{1/3}$ if
$\eper\neq 0$. On the other hand, if $\eper=0$, or if the critical
point is approached from the direction parallel to
$\vec{e}_{\parallel}$, one finds $\delta B\approx (-\epar)^{1/2}$.

\section{Correlation lengths}
It is established, in the context of the mean-field scenario of the
glass transition, that the cooperative dynamics are associated with
diverging correlation lengths near the dynamic glass transition
point. These lengths are predicted by the inhomogeneous version of the
MCT (IMCT) both for the $A_2$ and $A_3$ singularity
points~\cite{PhysRevE.58.3515,Physics.4.42,biroli2006inhomogeneous,nandi2014critical}.
In this section, we discuss the correlation lengths of the randomly
pinned FAM, using a point-to-set function, which is a quantity
introduced to analyze the spatial correlations in glassy
systems~\cite{bouchaud2004adam,franz2007analytic}.

\begin{figure}
 \onefigure[width=7cm]{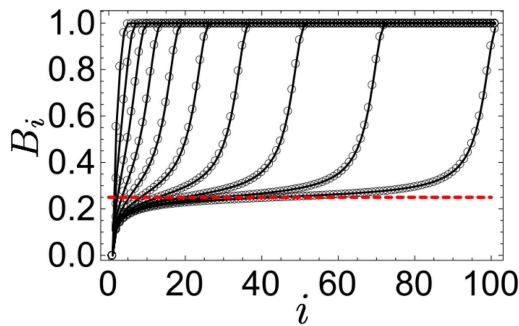} 
 \vspace{-0.5cm}
\caption{\small{ Dependence of $B_i$ on
the depth of the inner branch for $c=0$ and for various $p$'s.  We
define $(p_d-p)/p_d = 2^{-n}$ and the lines with circles
correspond to $n = 1, 2, \cdots, 10$ from left to right.  The dashed
horizontal line represents the threshold value $1/4$.}}
\label{020723_18Dec14}
\end{figure}
\begin{figure}
   \onefigure[width=7cm]{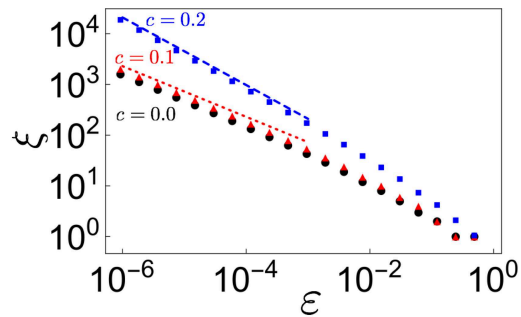} 
 \vspace{-0.5cm}
\caption{\small{$\xi$ as a
function of $\varepsilon =(p_d-p)/p_d$ for several $c$'s.  Circles
are for $c=0$, triangles for $c=0.1$, and squares are for
$c=0.2$.  Thin dotted and dashed lines represent
$\abs{\varepsilon}^{-1/2}$ and 
$\abs{\varepsilon}^{-2/3}$,
respectively.}}  \label{025039_18Dec14}
\end{figure}
\begin{figure}
\onefigure[width=7cm]{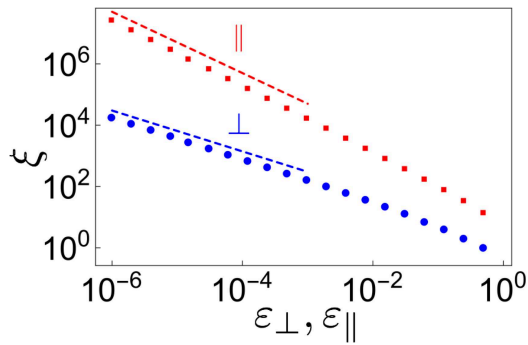} \caption{\small{$\xi$ near the
terminal critical point $(c_{s}=1/5$, $p_{s}=5/6)$ approached from
the parallel ($\perp$) and perpendicular ($\parallel$) directions.
Numerical results are shown in filled symbols.  Thin dashed lines
represent
$\abs{\varepsilon_{\perp}}^{-2/3}$ and $\abs{\varepsilon_{\parallel}}^{-1}$,
respectively.  }} \label{180201_26Dec14}
\end{figure}

The local structure of the Bethe lattice is Cayley-tree-like as shown in
Fig.~\ref{fig1}. We freeze the degrees of freedom of the spins at the
outermost branch $\sigma_0$ and set $B_{0}=0$. We consider how much the
effect of the frozen boundary penetrates down to the inner branches.
The value of $B_{i}$ down from the $0$-th node is written by the
recursive equation as
\begin{equation}
B_{i+1} = 
1-p + (1-c)p \left\{B_i^3+3B_i^2(1-B_i)\right\}.
\label{eq5}
\end{equation}
In Fig.~\ref{020723_18Dec14}, the numerical solution for $B_i$ of
eq.(\ref{eq5}) for $c=0$ is shown as a function of $i$ for several
values of $p$'s $(< p_d)$ in ergodic states.  $B_i$ for $i > 0$ remains
very small near the boundary but it relaxes to $1$ far away from the
boundary ($i\gg 1$).  As $p$ approaches to $p_d$ from below, the
distance over which the effect of the boundary penetrates grows.  We
shall define the correlation length $\xi$ as the $i$-th point at which
$B_{i}$ exceeds the threshold whose value is set to $B_d$, the solution
of eq.(\ref{eq2}) at $p=p_d$~\cite{schwarz2006onset}.  Numerical results
for $\xi$ for $c=0$, $0.1,$ and $0.2$ are shown as a function of the
distance from the transition point $\varepsilon =(p_d-p)/p_d$ in
Fig.~\ref{025039_18Dec14}.
It is clearly seen that $\xi$'s behave as $\abs{p_{d}-p}^{-1/2}$ for
$c=0$ and $0.1$~\cite{schwarz2006onset}.  But for $c=1/5$, one observes
$\xi\propto \abs{p_{d}-p}^{-2/3}$.  This asymptotic behavior of $\xi$
can be analytically explained by the linear analysis of
eq.(\ref{eq5}) around $p_d$.  Let us define the distance of
$B_i$ from the plateau value $B_d$ by $\delta B_i\equiv B_i-B_d$. We
assume that the variation of $B_{i}$ is small and can be represented
using a large number $l$ as $\delta B_{i+1}=(1-l^{-1})\delta B_i$.
Since $\delta B_{i}\approx (1-l^{-1})^{i}\approx e^{-i/l}$, $l$ can be
identified as $\xi$.  Substituting this expression back to the recursive
equation, eq.~(\ref{eq5}), we arrive at
\begin{equation}
\xi^{-1}\delta B_i = Q(c,p, B_i),
\end{equation}
where $Q(c,p, B_i)$ is defined by eq.(\ref{eq:Q-def}). Expanding $Q(c,p,
B_i)$ in terms of $\delta B_i$ around $B_d$ like eq.(\ref{eq:Q}), it is
straightforward to show that $\xi$ behaves as $\xi \propto
\varepsilon^{-1/2}$ when $c < 1/5$.  In the vicinity of the terminal
critical point $(c_s, p_s)=(1/5, 5/6)$, the behavior of $\xi$
qualitatively changes. If 
$\eper\neq 0$, we have $\xi \propto
\eper^{-2/3}$, thus reproducing the results of
Fig.~\ref{025039_18Dec14}.  If
$\eper=0$, one finds $\xi \propto
\epar^{-1}$. The direction dependence of the growth of $\xi$ is shown
in Fig.~\ref{180201_26Dec14} together with the direct numerical results.
It is known that the exponents for the length scales of the model
systems on the Bethe lattices are twice as large as the corresponding
values of the mean-field
theory~\cite{grimmett2010percolation,schwarz2006onset,franz2011analytical}.
Therefore, the exponent of the diverging length $\xi \sim
\varepsilon^{-\nu}$ obtained above should correspond to the value of the
mean-field theory of $\nu = 1/4$ instead of $1/2$ for $c<1/5$.
Likewise, near the terminal critical point, the exponents should be
$\nu_{\perp}=1/3$ for $\eper\neq 0$ and $\nu_{\parallel}=1/2$ for
$\eper= 0$, respectively.  These results are equivalent with those
obtained from the IMCT~\cite{biroli2006inhomogeneous,nandi2014critical}.
\section{Conclusion}
In this letter, the stationary properties of the randomly pinned
FAM on the Bethe lattice have been discussed.  The order
parameter, $\phi$, or the probability that the spin flips, $B$, exhibits
analogous behavior as the nonergodic parameter of the mean-field models
of the glass transition in the presence of the randomly pinned
spins/atoms at their dynamic transition
lines~\cite{cammarota2013random,cammarota2012ideal,PhysRevLett.94.065703}.
As the fraction of the pinned spins, $c$, increases, the $A_2$ singular
point $p_d(c)$, at which the order parameter discontinuously jumps,
decreases (or $T_d(c)$ increases) and terminates at a larger but finite
$c=c_s$, where the transition point becomes the higher order singular
point of type $A_3$, which is fully consistent with the dynamic
transition lines predicted by the MCT for randomly pinned systems.  The
concept of the point-to-set correlation has been employed to evaluate
the correlation length, $\xi$, around this transition line,
$p_d(c)$~\cite{bouchaud2004adam,franz2011analytical}. Aside from the
trivial factor of $2$, the exponent, $\nu$, of the correlation lengths,
$\xi \sim\varepsilon^{-\nu}$, is consistent with that obtained from the
inhomogeneous version of the MCT \cite{nandi2014critical}.  The exponent
at the $A_3$ singularity point is $\nu_{\perp} =1/3$ and
$\nu_{\parallel} =1/2$, depending on the directions to approach the
critical point.  This behavior is akin to that of the ferromagnetic
transition in the Landau theory ($\nu=1/2$ when the magnetic field $h=0$
and $T \rightarrow T_c$ and $\nu=1/3$ when $T=T_c$ and $h \rightarrow
0$~\cite{nandi2014critical}). Note that the parallel
direction is more special than the perpendicular direction, because the
critical behavior along the former is observed only when
$\varepsilon_{\perp}= 0$, whereas the later is generic in a sense that
the critical behavior is observed even if $\varepsilon_{\perp}\neq 0$.
We should emphasize, however, that the analysis of the order parameter
as studied here can not establish the universality class of the
transition.  It is known that the transition of the mean-field
random-field Ising model (RFIM) is also characterized by the same
exponents as those of the ferromagnetic transition in the Landau theory
but its universality class is distinct~\cite{nishimori2010elements}.
The analysis of the fluctuations such as the higher order moments is
necessary to characterize the nature of the transition.  Indeed, it has
been demonstrated that from the dynamic scaling of the model discussed
here at $c=0$, the model belongs to the RFIM universality
class\cite{franz2013finite}.  Besides, the randomly pinned glass model
is also argued to belong to the same universality
class~\cite{biroli2014random}.

In this letter, we only analyzed the stationary properties of the
randomly pinned FAM.  It is known that the slow relaxation dynamics of
the FAM on the Bethe lattice for $c=0$ is qualitatively similar to that
of the MCT for the bulk supercooled liquid or mean-field spin glass
models\cite{sellitto2005facilitated}.  On the other hand, the MCT and
IMCT predict that the slow dynamics near the $A_3$ singularity point
is distinct from that at the $A_2$ point, and the system exhibits
anomalous dynamics such as the logarithmic relaxation and milder growth
of the correlated
fluctuations\cite{gotze2008complex,nandi2014critical}. Studies in this
direction are left for future works.

\acknowledgments We thank G. Biroli for comments.  We acknowledge
KAKENHI No. 24340098, 25103005, 25000002, and the JSPS Core-to-Core program.
H. I. was supported by
Program for Leading Graduate Schools ``Integrative Graduate Education
and Research in Green Natural Sciences'', MEXT, Japan.

\end{document}